# Spin-Dependent Electron Transport through Bacterial Cell Surface Multiheme Electron Conduits

Suryakant Mishra,[†#] Sahand Pirbadian,[‡#] Amit Kumar Mondal,[†] Mohamed Y. El-Naggar,[*‡‖§] and Ron Naaman[*†]

† Department of Chemical and Biological Physics, Weizmann Institute of Science, Rehovot 76100, Israel

‡ Department of Physics and Astronomy, University of Southern California, Los Angeles, California 90089, United States

‖ Department of Chemistry, University of Southern California, Los Angeles, California 90089, United States

§ Department of Biological Sciences, University of Southern California, Los Angeles, California 91030, United States

*Supporting Information Placeholder*

**ABSTRACT:** Multiheme cytochromes, located on the bacterial cell surface, function as long-distance (> 10 nm) electron conduits linking intracellular reactions to external surfaces. This extracellular electron transfer process, which allows microorganisms to gain energy by respiring solid redox-active minerals, also facilitates the wiring of cells to electrodes. While recent studies suggested that a chiral induced spin selectivity effect is linked to efficient electron transmission through biomolecules, this phenomenon has not been investigated in the extracellular electron conduits. Using magnetic conductive probe atomic force microscopy, Hall voltage measurements, and spin-dependent electrochemistry of the decaheme cytochromes MtrF and OmcA from the metal-reducing bacterium *Shewanella oneidensis* MR-1, we show that electron transport through these extracellular conduits is spin-selective. Our study has implications for understanding how spin-dependent interactions and magnetic fields may control electron transport across biotic-abiotic interfaces in both natural and biotechnological systems.

Electron flow dictates all biological energy conversion strategies.[1,2] In the case of respiration, cells harvest energy by controlling electron flow from electron donors (fuels) to terminal electron acceptors (oxidants) through a chain of reduction-oxidation (redox) cofactors. Some microorganisms (including metal-reducing bacteria) can also extend this electron transport chain to terminal acceptors *outside* the cells, allowing anaerobic respiration of solid minerals in the absence of soluble oxidants (e.g. $O_2$) that enter the cells.[3] This extracellular electron transfer (EET) strategy also facilitates the ‘wiring’ of cells to solid-state electrodes in technologies such as microbial fuel cells, electrosynthesis and bioelectronics.[4–7]

The metal-reducing bacterium *Shewanella oneidensis* MR-1 expresses a network of multiheme cytochromes (MHCs), known as the Mtr-Omc pathway, to accomplish EET across the biotic-abiotic interface.[8] As part of this pathway, decaheme cytochromes located on the cell surface (MtrC, MtrF, OmcA), can transmit electrons from periplasmic redox partners to the extracellular space.[9–11] Measurements[12–14] and quantum/molecular simulations[8,15] revealed rapid electron hopping rates through the *S. oneidensis* multiheme conduits, sufficiently high to meet the cellular EET rate.[16] These cytochromes can also facilitate long-distance (micrometer scale) redox conduction along cellular membranes.[17] Rapid electron flux ($10^5$ $s^{-1}$) through the solvated decaheme cytochromes is thought to arise from the packing of hemes into molecular wire-like chains, presence of cysteine linkages that enhance electronic couplings, and careful control of the redox potential landscape.[8,15,18,19] Solid-state (vacuum) measurements in monolayer junctions of the MHCs also revealed remarkable temperature-independent electronic conduction (0.3 A $cm^{-2}$ at 50 mV for MtrF), on par with conjugated organics, suggesting a heme-assisted coherent tunneling mechanism.[20]

An additional factor that may enhance the electron transport efficiency in biological systems has recently been observed: chiral induced spin selectivity (CISS), an effect that couples the electron’s spin to its linear momentum in a chiral potential, including nucleic acids and proteins.[21,22] This property enhances the transmission probability of one preferred spin, dependent on the chirality of the molecule, and suppresses backscattering.

Given the observations of efficient electron flux in bacterial MHCs, we hypothesized that electron transport through these proteins could accompany spin selectivity. CISS in MHCs could potentially give rise to spin effects in the biotic-abiotic interaction between cells and solid phase electron donors/acceptors, especially those with magnetic properties, or in electron-exchange processes involving other chiral biomolecules, such as electron shuttling or interspecies electron transfer. In addition, CISS in MHCs may provide a basis for understanding the reported magnetic field effect on the performance of microbial fuel cells.[23–26], where it has been observed that static magnetic fields of specific magnitudes and directions can improve overall cell-anode EET. The latter observations have so far been tentatively assigned to oxidative stress or magnetohydrodynamic effects, but the role of spin has not been investigated. We note that possible roles of CISS in various biological processes have been reviewed elsewhere.[21]

Here, we investigated and confirmed the role of spin in electron transport through the *S. oneidensis* MR-1 outer membrane decaheme cytochromes MtrF and OmcA. To monitor electron transport, extent of spin polarization in the transferred electrons, and spin-dependent polarizability in the proteins, we applied various experimental techniques previously used to demonstrate CISS in DNA, oligopeptides, and chiral polymers: solvent-free magnetic conductive probe atomic force microscopy (mCP-AFM), Hall voltage measurements along with spin-dependent electrochemistry of the proteins in solution.[27–31]

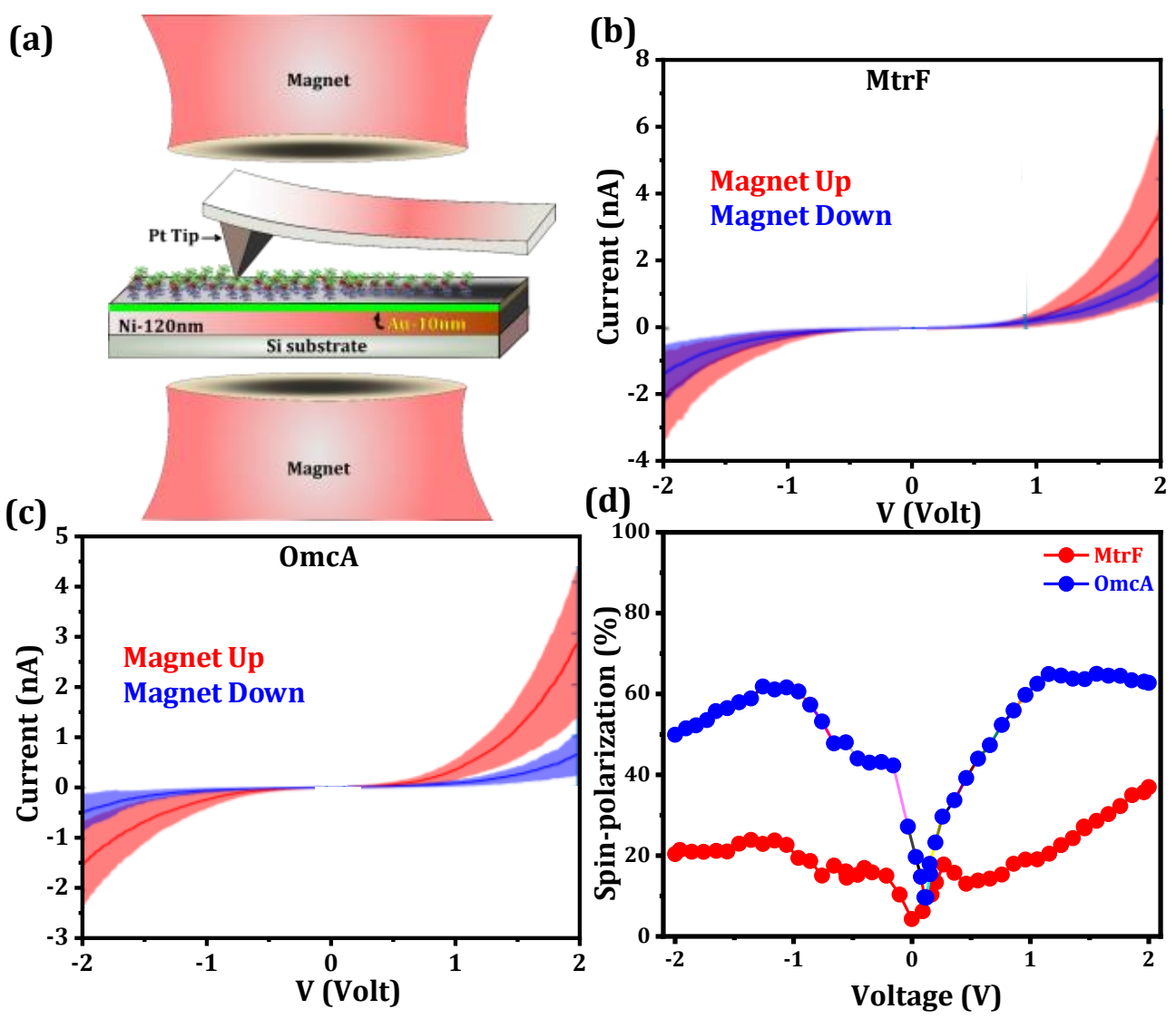

**Figure 1.** Spin-dependent conduction study of MtrF and OmcA by mCP-AFM. **(a)** Scheme of the measurement, **(b & c)** *I-V* plots of MtrF and OmcA, respectively where Ni film magnetized with the north pole pointing up (red) or down (blue). **(d)** The corresponding percentage of spin-polarization (SP) {[($I_{up}$ – $I_{down}$)/($I_{up}$ + $I_{down}$)]×100} for MtrF and OmcA, respectively. Here $I_{up}$ and $I_{down}$ are the currents with magnetic north pole up and down, respectively. (Note: panels **b)** and **c)**, width of the lines represents the standard deviation of the measurements.)

Using mCP-AFM, we measured electron transmission through solvent-free MtrF and OmcA adsorbed on a ferromagnetic Ni, 120nm thick substrate coated with a thin (10 nm) Au layer. MtrF and OmcA were effectively immobilized on the surface through covalent thiol bonds with Au as a result of a recombinant tetra-cysteine tag at the C-terminus of the proteins, as described in previous scanning probe studies[14] and confirmed here (Figure S2). Nonmagnetic (Pt) tips functioned as the top electrodes and conduction was measured with magnetic fields pointing either with the north pole UP or DOWN using a permanent magnet that determines the spin alignment in the Ni bottom substrate.[28,32] Current-voltage (I-V) spectra were acquired from multiple points on each of the monolayers, revealing a magnitude and voltage dependence consistent with previous tunneling spectroscopy measurements of both proteins.[12,14]

Figure 1 shows clear spin selectivity in both proteins, with higher conductivity when the magnetic field is pointing ‘UP’ compared to ‘DOWN’. The extent of SP at a given voltage can be quantified using the ratio ($I_{UP}$-$I_{DOWN}$)/($I_{UP}$+$I_{DOWN}$), where $I_{UP}$ and $I_{DOWN}$ are the currents associated with the two different magnetic field directions. As can be seen in Figure 1d, OmcA displayed the higher SP (63 ± 2%) compared to MtrF (37 ± 3%) at 2.0 V bias.

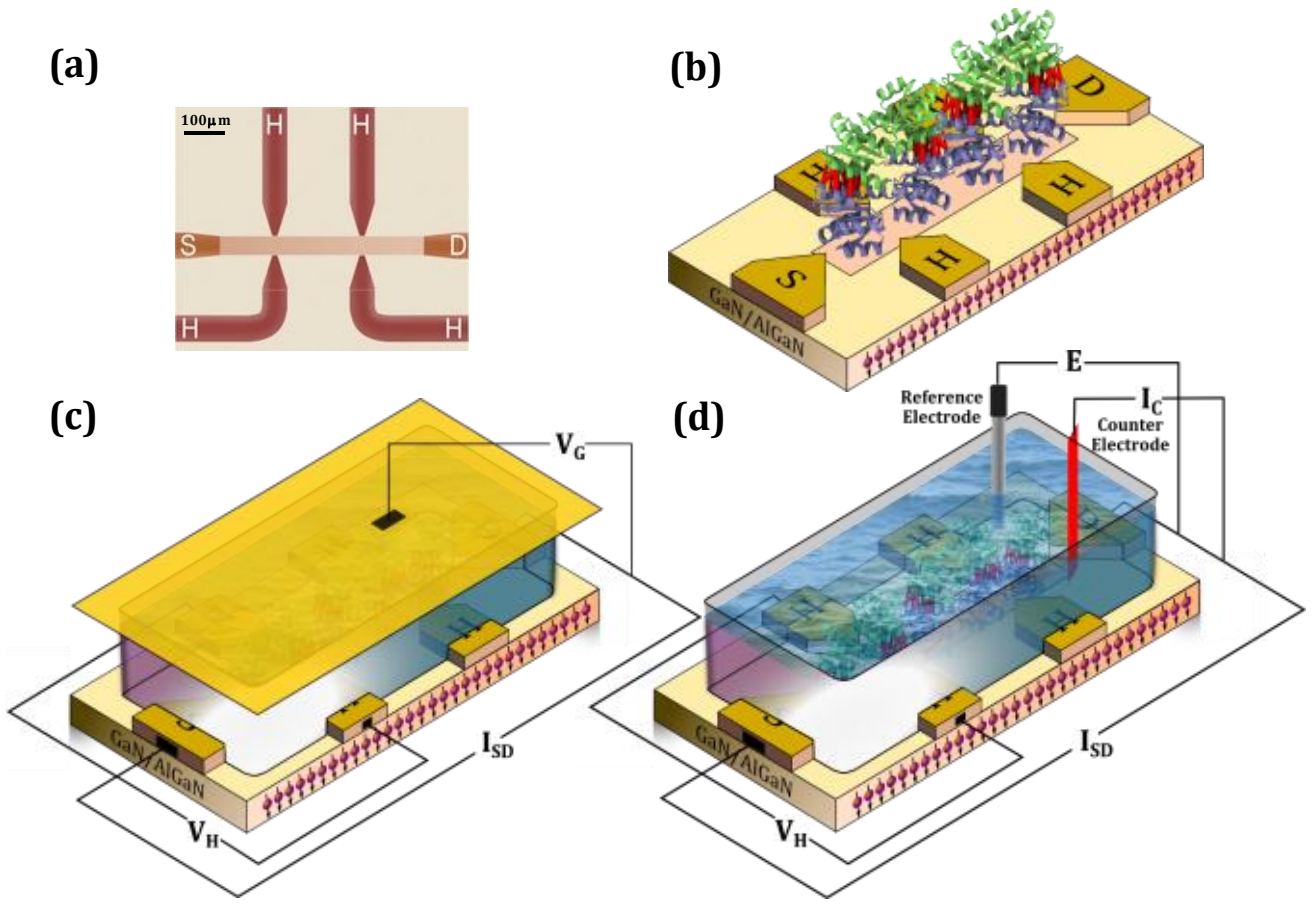

**Figure 2. (a)** Optical microscopic image of the Hall device patterned on GaN/AlGaN substrate. **(b)** A scheme of the Hall device on which a monolayer of the protein is adsorbed. **(c)** A scheme of the setup used for measuring spin polarization. A Hall device coated with monolayer of proteins is covered by buffer electrolyte with top gate electrode insulated from the solution. **(d)** Spin-dependent electrochemistry setup where Hall device used as the working electrode measures the faradaic current flows through the protein monolayer and Hall potential.

We also measured the Hall voltage resulting from the spin polarizability that accompanies charge polarization across MtrF and OmcA in solution (see SI for details). The measurement system (Figure 2) is based on Au-coated (5 nm film) Hall device patterned on a GaN/AlGaN two-dimensional electron gas structure.[29,30] In addition to allowing thiol-binding, the Au film stabilizes the potential on the surface by eliminating surface states.[30] With a constant driven source-drain current, a voltage is applied between a top gate electrode and the device on which the proteins are placed. The gate voltage generates an electric field that induces charge polarization perpendicular to the protein monolayer. If this charge polarization is accompanied by spin polarization, a magnetic field is created and a Hall voltage is measured across the lateral Hall probes (Figure 2). Prior to Hall voltage measurements, the attachment of the MtrF and OmcA protein monolayers were confirmed with liquid tapping mode atomic force microscopy and polarization modulation-infrared reflection-absorption spectroscopy (Figure S2 and S3).

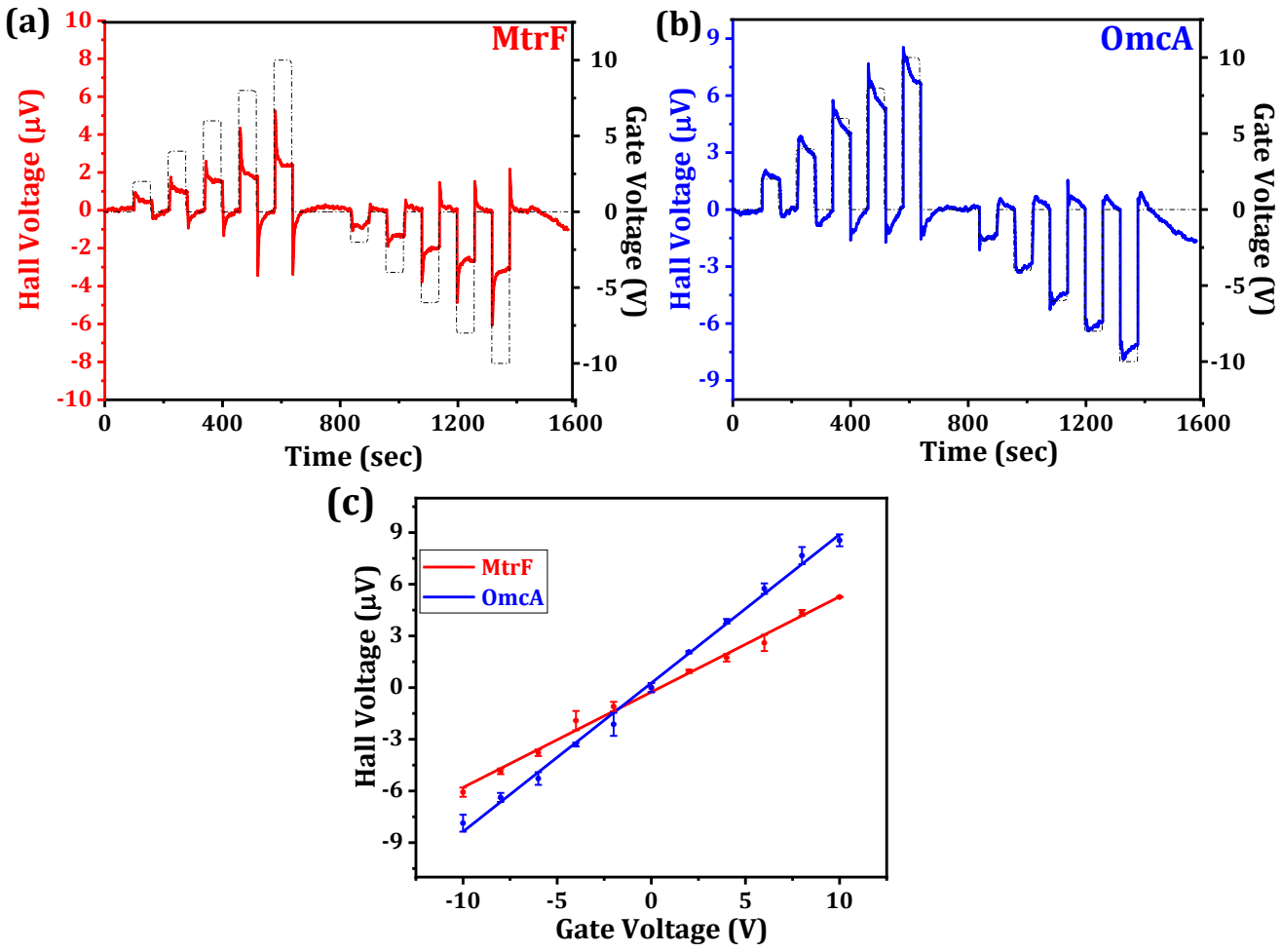

**Figure 3.** The spin polarizability measured as a function of the potential applied (dotted black line) on the top gate of gold (Fig. 2C) for **(a)** device coated with MtrF and **(b)** device coated with OmcA. **(c)** The Hall signal as a function of the gate voltage applied for devices coated with MtrF (red) and OmcA (blue). A linear response is observed and OmcA having higher spin polarizability compared to MtrF.

Figure 3(a, b) shows the Hall voltages observed in response to gate voltage steps of different magnitudes and signs for both MtrF and OmcA. This data confirm that the spin polarization indeed accompanies the field-induced charge polarization in both MHCs. The Hall signals scale linearly with the gate voltage (Figure 3c) and, consistent with the mCP-AFM measurements, the spin polarization is larger for OmcA as compared to MtrF. By comparing to a separate calibration of the Hall devices using external magnetic fields (see Figure S4 in SI), the OmcA Hall signal at 10 V gate voltage corresponds to a magnetic field of about 200 Gauss. To verify the importance of the secondary/tertiary structure in the observed spin polarization effect, the proteins were denatured at 80 ˚C (see SI for details), after which no spin polarization was observed in response to gate voltage (Figure S5).

It is interesting to consider the possible reasons leading to higher spin polarization in OmcA relative to MtrF. The two MHCs have comparable conductivities (Figure 1), so it is unlikely that the difference results from higher overall electron transmission. Another factor may be protein size, since the field-induced electric dipole moment depends on the size of the protein. OmcA (83 kDa) is moderately bulkier than MtrF (74 kDa).[8] However, a comparison of the X-ray structures shows similar overall dimensions, particularly along the charge carrier heme chains, that define the cross configuration common to both proteins.[33,34] Differences in overall size are therefore a less likely explanation for the significant difference in spin selectivity. A comparison of the secondary structures, however, offers clues. For example, α-helices serve as primary scaffolds for hemes in both proteins, but OmcA has a significantly higher helical secondary structure (18%) than MtrF (11%) when compared using the DSSP tool.[35] Figure S6 highlights the increased helical content in the heme-containing domains II and IV of OmcA compared to MtrF. We therefore hypothesize that the difference in the secondary structure surrounding the electron carrying heme chains plays an important role in determining the extent of the spin selectivity.

In a third experimental approach, we performed spin-dependent electrochemistry as previously applied to DNA and oligopeptides.[30,31] Here, measurements are performed in 3-electrode electrochemical cells with the Hall device serving as the working electrode. While performing cyclic voltammetry (CV), the Hall potential is monitored simultaneously (Figure S7a&b). It is interesting to note that the electrochemical potentials of MtrF and OmcA are shifted relative to previous reports, [33,36,37] which we attribute to the immobilization strategy and the use of bare thin gold electrodes, rather than adsorption on graphite or self-assembled monolayers, since the immobilization procedure can influence the measured redox properties.[37] In addition to the reductive and oxidative electrochemical signatures observed in the CVs of MtrF and OmcA, we observed a simultaneous Hall signal reflecting spin selectivity associated with the electron transfer through both proteins (Figure S7b). It is important to note, however, that the CV redox peaks are not reflected in the Hall signals. This finding is consistent with a previous spin-dependent electrochemistry study where chiral oligopeptides interact with a redox probe,[31] and demonstrate that spin selectivity rises from electron conduction through the protein rather than redox processes of the hemes themselves.

Like the mCP-AFM (Figure 1) and field-induced polarization measurements (Figure 3), higher spin selectivity was observed for OmcA, compared to MtrF, in the electrochemical measurements (Figure S7a&b). Denaturation resulted in significant decrease of electrochemical current and corresponding order of magnitude reduction in the Hall signal (Figure S7c-f), again confirming the role of the protein's structure in dictating the spin selectivity process associated with electron transmission.

While the theoretical basis of CISS is not fully worked out, it is currently understood as a *dynamic* phenomenon where the chiral environment couples the electron spin direction and velocity so that each conduction direction has a preferred spin alignment.[38] In this sense, CISS is associated with electron transmission through the molecule, rather than spin state of the charge carriers themselves (e.g. the heme redox centers). Previous EPR measurements of the MHCs reveal that the oxidized hemes are in a low spin ($S=\frac{1}{2}$) state while the reduced hemes are EPR-silent ($S=0$).[39] It will be interesting in the future, given the availability of structures and electronic structure calculations,[15] to consider whether CISS interacts with the spin states of the hemes or transient high spin intermediates in the proteins.

The spin selectivity observed in the extracellular bacterial cytochromes may have interesting implications for controlling electron transfer across the biotic-abiotic interface. It was recently proposed that such spin selectivity may place constraints on the ability to interact with other chiral molecules.[29] In the case of EET conduits, this effect may lead to selectivity in interactions with electron exchange partners, including soluble redox shuttles such as flavins or neighboring electron conduits of other cells. We also speculate that spin selectivity may play a role in controlling interactions with external electron accepting minerals, such as certain iron oxides, that have magnetic properties. Finally, spin polarization offers a concrete mechanism that may impact our understanding of multiple recent reports[23–26] describing magnetic field enhancements of EET in microbial fuel cells.

To summarize, we have found that electron flow in the *Shewanella oneidensis* MR-1 cell surface decaheme cytochromes is spin selective. This observation opens up an additional degree of freedom, based on electron spin, for controlling charge transport across biotic-abiotic interfaces in both natural and biotechnological applications.

## ASSOCIATED CONTENT

**Supporting Information**.

The Supporting Information is available free of charge on the ACS Publications website.

Experimental details (PDF).

## AUTHOR INFORMATION

**Corresponding Authors**

* mnaggar@usc.edu, ron.naaman@weizmann.ac.il

**Author Contributions**

# These authors contributed equally.

**Notes**

The authors declare no competing financial interests.

## ACKNOWLEDGMENT

M.Y.E.-N. and S.P. acknowledge support by the U.S. Office of Naval Research Multidisciplinary University Research Initiative Grant No. N00014-18-1-2632. We are grateful to Prof. Liang Shi (China University of Geoscience in Wuhan) for supplying the MtrF and OmcA constructs. RN acknowledges the support of the Templeton Foundation and of the Israel Science Foundation.

TOC

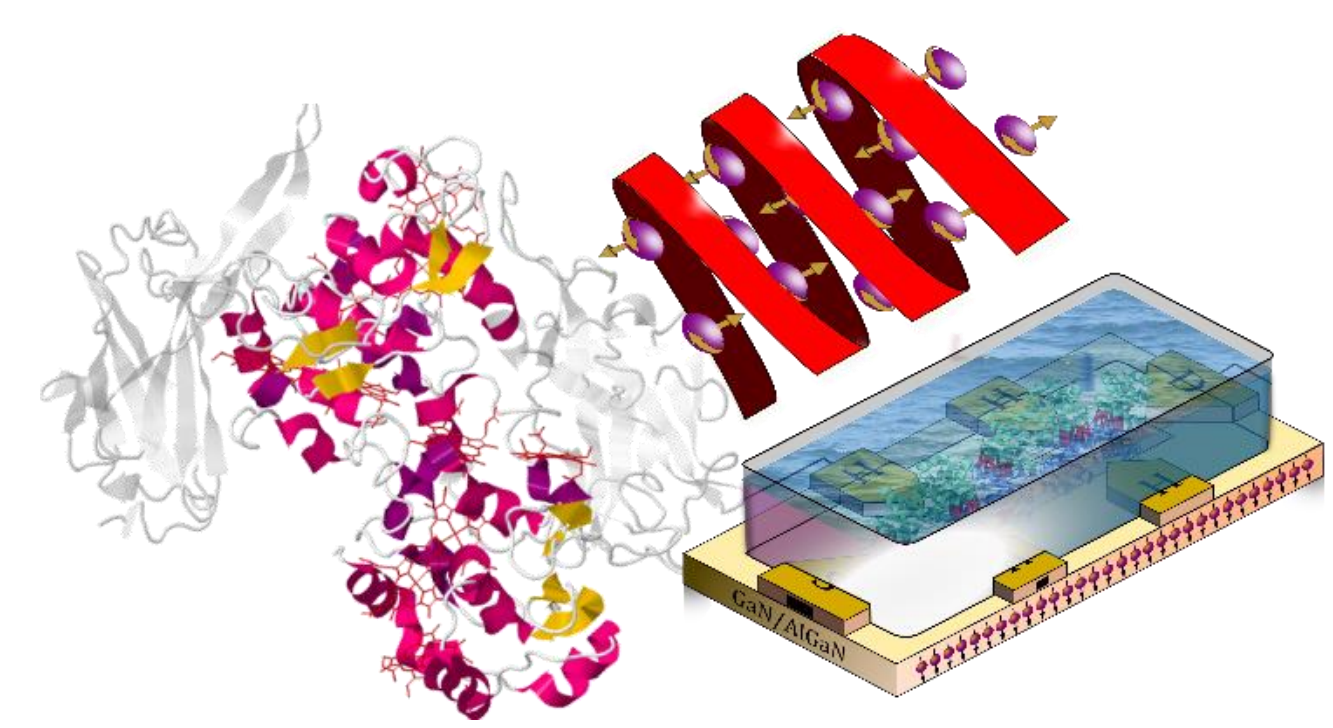